# A Graph Neural Network for Global Daily Fire Radiative Power Prediction at Medium-Range Lead Times

Li Zhang[1, 2], Jun Wang[3], Isidora Jankov[2], Yongxin Liu[4], Gonzalo A. Ferrada[1, 2], Ravan Ahmadov[2], Ligia Bernardet[2], Haonan Chen[4], Shobha Kondragunta[5]

[1]*CIRES, University of Colorado Boulder, Boulder, CO, US;*

[2]*Global Systems Laboratory, OAR/NOAA, Boulder, CO, US;*

[3]*NOAA/NWS/OMD, College Park, MD, US;*

[4]*CIRA, Colorado State University, Fort Collin, CO, US;*

[5]*NOAA/NESDIS Center for Satellite Applications and Research, College Park, MD, US;*

This Work has been submitted to *Artificial Intelligence for the Earth Systems*. Copyright in this Work may be transferred without further notice.

October 7, 2026

*Corresponding author*: Li Zhang (kate.zhang@noaa.gov)

## Abstract

Skillful prediction of biomass-burning activity several days in advance is important for air-quality forecasting and aerosol prediction. Two operational constraints motivate this work. First, the GBBEPx satellite fire radiative power (FRP) product used to initialize NOAA's GEFS-Aerosols is available with about a 1.5-day latency, so each forecast cycle relies on the most recently available, but already outdated, fire observations. Second, these fire inputs are then held fixed throughout the subsequent 5-day operational forecast, or 7 days in the GSL experimental system, effectively assuming no evolution in fire activity. We develop a data-driven model that predicts global FRP one to seven days ahead from the most recent available observations. The model adapts a spatiotemporal graph neural network using reanalysis meteorology, land-cover and vegetation information, recent fire history, and GBBEPx FRP as the training target. It is trained on 2020–2022 data and evaluated for 2023–2024. The model reproduces the global seasonal cycle and substantially outperforms persistence. At 0.1° resolution, mean squared error is reduced by 32% at one-day lead and 43% at seven days in 2023, and by 24% and 40% in 2024. At 1° resolution, the critical success index ranges from 0.32 to 0.60. Detection skill declines only modestly with lead time, whereas intensity skill degrades more rapidly. Large fires are detected reliably, but their radiative power is systematically underestimated. These results demonstrate useful predictability of fire activity several days ahead and identify intensity calibration and small-fire placement as the main remaining challenges before predicted FRP can support operational aerosol forecasts.

**Significance Statement.** Air-quality forecasts depend on knowing where fires are burning, but the satellite observations that supply this information arrive about a day and a half late and are then held unchanged for the rest of the NOAA operational forecast — effectively assuming fires do not change for up to a week. We show that a machine-learning model can instead forecast daily global fire radiative power out to seven days, reducing error relative to that assumption by a third or more, with the largest gains at the end of the forecast period where the assumption is weakest. The model locates large fires reliably but under-predicts their intensity, which is the main obstacle before these forecasts can drive operational smoke predictions.

## 1. Introduction

Landscape fires are a major component of the Earth system, exerting strong control on atmospheric composition, air quality, the carbon cycle, and regional radiative forcing (Bowman et al. 2009; Andela et al. 2017). Biomass burning is one of the largest global sources of carbonaceous aerosols and trace gases (Andreae and Merlet 2001; Bond et al. 2013; Andreae 2019), and smoke plumes from intense fire events can degrade air quality far downwind of the source region, with documented impacts on human health and visibility (Johnston et al. 2012; Reid et al. 2016). Within numerical weather and Earth system prediction frameworks, fire emissions are an essential lower boundary condition for aerosol and chemistry components, and errors in the magnitude or location of emitted smoke propagate directly into predicted aerosol optical depth, radiative fluxes, and aerosol–cloud interactions (Grell et al. 2011; Kaiser et al. 2012). Consequently, predicting fire evolution is fundamental to extending smoke and atmospheric-composition forecast skill beyond the first few days.

Fire radiative power (FRP), the radiant energy released per unit time by actively burning vegetation, is a physically based and globally observable proxy for fire intensity and combustion rate (Kaufman et al. 1998; Wooster et al. 2005). FRP retrievals from polar-orbiting and geostationary satellites underpin operational biomass-burning emission inventories that drive global aerosol forecasts (Ichoku and Kaufman 2005; Kaiser et al. 2012; Zhang et al. 2012). Predicting future FRP therefore provides a direct basis for predictive fire emissions.

A major limitation of current smoke-prediction systems is that future fire emissions are generally prescribed from recently observed fire activity rather than predicted dynamically. NASA GEOS-CF persists QFED biomass-burning emissions through its five-day forecasts (Darmenov and da Silva 2015; Keller et al. 2021), while ECMWF CAMS similarly holds GFAS emissions fixed during five-day integrations (Kaiser et al. 2012; Di Giuseppe et al. 2017). In NOAA's operational global aerosol prediction system, GEFS-Aerosols, GBBEPx provides FRP-based fire location and intensity for carbonaceous aerosol and $SO_2$ emissions, together with an online plume-rise module. Because GBBEPx has a latency of about 1.5 days, each cycle begins with emissions already about 1.5 days old and then persists them through the forecast (Zhang et al. 2022, 2026). Regional North American smoke forecasting systems have similarly relied on recently observed fire activity (Rolph et al. 2009; Larkin et al. 2009; Ahmadov et al. 2017). Such approaches cannot anticipate new ignitions, rapid fire growth, changes in intensity, geographic spread, or extinction. Persistence therefore becomes increasingly problematic with lead time. Di Giuseppe et al. (2017) improved CAMS aerosol forecasts by modulating persisted emissions using daily changes in the fire weather index (FWI), but such approaches still rely on prescribed empirical relationships rather than directly predicting fire evolution from environmental conditions.

Despite the limitations of persistence, fire activity is not entirely unpredictable. Its evolution is governed by interactions among fire weather—including temperature, humidity, wind, and precipitation—fuel availability and dryness, vegetation state, antecedent moisture, and ignition processes (Bradstock 2010; Abatzoglou and Williams 2016). These relationships are strongly nonlinear, spatially heterogeneous, and characterized by extreme value behavior, with a small number of large events dominating total emitted energy (Abatzoglou and Williams 2016; Archibald et al. 2013). Recent data-driven studies demonstrate that this environmental information contains useful predictive information. Random-forest models have been used to forecast daily fire radiative energy over the western United States and substantially outperform persistence at one- and two-day lead times (Thapa et al. 2024). Machine-learning fire spread systems such as Fire Intensity and spRead forecAst (FIRA) have generated FRP fields for regional air-quality applications (Hung et al. 2025), the Hourly Wildfire Potential (HWP) index uses High-Resolution Rapid Refresh (HRRR) meteorology to predict sub-daily variability in the activity of existing fires over the contiguous United States at lead times of 1–2 days (James et al. 2025). At the global scale, ECMWF's data-driven Probability of Fire system predicts the daily probability of satellite-detected active fires at about 9-km resolution up to about 10 days ahead (McNorton et al. 2024; Di Giuseppe et al. 2025), but it forecasts fire occurrence rather than fire intensity. Other existing approaches remain largely regional, short-range, or focused on fire danger, occurrence, spread, or related proxies rather than a global gridded FRP field directly usable as an emissions boundary condition.

The strong nonlinear dependence of fire activity on spatially and temporally evolving environmental conditions makes it well suited to data-driven prediction. Graph neural networks (GNNs) are particularly attractive because they represent spatial interactions through message passing while operating efficiently over global domains. Recent advances in data-driven weather prediction demonstrate that GNNs can produce skillful spatiotemporal forecasts of geophysical fields. GraphCast, for example, represents the atmosphere as a graph and advances its state through learned message passing (Lam et al. 2023). These same principles—spatial information exchange combined with temporal encoding of recent environmental conditions—provide a natural framework for fire prediction.

Predicting FRP also poses statistical challenges: the field is extremely sparse, and its nonzero values are heavy-tailed and span several orders of magnitude, so a model trained with a mean-squared-error objective tends toward near-zero values and misses rare large fires. This motivates a formulation that separates fire occurrence from intensity and emphasizes large fires.To address these challenges, this study develops and evaluates a GNN-based system for global daily FRP prediction at lead times of one to seven days, verified against two years withheld from training, 2023 and 2024. The contributions of this work are fourfold. First, we develop a GNN framework that predicts both fire occurrence and fire intensity using a hurdle-type prediction head and a composite loss function designed for sparse, heavy-tailed FRP observations. Second, we introduce a five-domain training strategy with domain-specific settings that accommodates strong contrasts among global fire regimes, including boreal and temperate crown fires, tropical savanna fires, and equatorial peatland fires. Third, we introduce a physically motivated, climatology-constrained per-pixel intensity cap that anchors predicted fire intensity to the observed historical FRP distribution at inference time. Fourth, we assess the forecast directly against the persistence assumption it is intended to replace—holding the most recent available observed FRP fixed at all lead times—and quantify the margin at the 0.1° resolution on which emission fields are consumed. Because persistence is the practical operational baseline in the absence of a global multi-day FRP forecast, skill relative to it directly measures potential operational benefit.

The remainder of the paper is organized as follows. Section 2 describes the observational and predictor datasets. Section 3 presents the model architecture, the domain-based training strategy, the loss function, and the climatology-constrained intensity cap. Section 4 describes the verification framework. Section 5 presents the results, and Section 6 discusses their interpretation, summarizes the conclusions, and outlines future work.

## 2. Data

### *2.1. Fire Radiative Power Target*

The prediction target is daily FRP from the NOAA operational Blended Global Biomass Burning Emissions Product version 4 (GBBEPx v4), which has been in real-time operational use since 2019. GBBEPx v4 combines fire detections and FRP retrievals from multiple polar-orbiting satellite instruments into a gridded, gap-filled global product at 0.1° × 0.1° horizontal resolution and provides daily global emissions of particulate matter with aerodynamic diameter ≤ 2.5 μm ($PM_{2.5}$), black carbon (BC), organic carbon (OC), carbon monoxide (CO), carbon dioxide ($CO_2$), and sulfur dioxide ($SO_2$) (Zhang and Kondragunta 2019). Its spatial continuity, high resolution, and operational use make it an appropriate training and verification target. The daily mean FRP field is used as the predictand throughout this study.

*2.2. Meteorological Predictors*

Meteorological predictors are derived from the fifth-generation ECMWF atmospheric reanalysis (ERA5; Hersbach et al. 2020). The native 6-hourly reanalysis fields are aggregated to daily statistics and bilinearly interpolated onto the common 0.1° × 0.1° grid of the GBBEPx v4 target, so that all predictors and the predictand share a single spatial grid. Most predictors are represented by their daily mean. For 700-hPa relative humidity and lower-tropospheric lapse rate, daily extrema are also retained because within-day extremes are relevant to fire growth. Precipitation is represented by daily accumulation, 10-m wind gust by the daily 90th percentile, and 500-hPa vertical velocity by the daily minimum, corresponding to the strongest ascent. The resulting predictor set therefore captures both mean fire-weather conditions and relevant within-day extremes.

The 27 meteorological predictors cover near-surface thermodynamics, surface and lower-tropospheric winds, moisture and precipitation, boundary-layer and stability structure, and surface radiation (See Table 1). Antecedent precipitation is encoded as an exponentially weighted accumulation,

$$P_decay(t) = \alpha \cdot P_decay(t - 1) + P(t), \quad \alpha = 0.7,$$

which provides a simple representation of antecedent fuel-moisture memory. The complete set of meteorological predictors, with their source reanalysis fields and units, is listed in Table 1.

| **Feature name** | **Description** | **ERA5 source field** | **Aggregation** | **Units** |
|---|---|---|---|---|
| *t2m_mean* | 2-m air temperature | 2m_temperature | Daily mean | K |
| *rh_mean* | 2-m relative humidity | 2m_temperature, 2m_dewpoint_temperature (derived) | Daily mean | % |
| *vpd_mean* | Vapor-pressure deficit (2 m) | 2m_temperature, 2m_dewpoint_temperature (derived) | Daily mean | hPa |
| *10u_mean* | 10-m zonal (eastward) wind | 10m_u_component_of_wind | Daily mean | m $s^{-1}$ |
| *10v_mean* | 10-m meridional (northward) wind | 10m_v_component_of_wind | Daily mean | m $s^{-1}$ |
| *gust_p90* | 10-m wind gust | 10m_wind_gust_since_previous_post_processing | Daily 90th pct | m $s^{-1}$ |
| *precip_sum* | Total precipitation | total_precipitation | Daily accumulation | mm $day^{-1}$ |
| *precip_decay* | Exponentially decaying antecedent precipitation ($\alpha$ = 0.7) | total_precipitation (derived) | Derived, daily | mm |
| *pblh_mean* | Planetary boundary-layer height | boundary_layer_height | Daily mean | m |
| *swdown_mean* | Surface downward shortwave radiation | surface_solar_radiation_downwards | Daily mean | W $m^{-2}$ |
| *tcwv_mean* | Total column water vapor | total_column_water_vapor | Daily mean | kg $m^{-2}$ |
| *soilm_rootzone* | Root-zone soil moisture (mean of layers 1–3, 0–100 cm) | volumetric_soil_water_layer_1–3 | Daily mean | $m^3$ $m^{-3}$ |
| *rh700_mean* | Relative humidity at 700 hPa | relative_humidity @ 700 hPa | Daily mean | % |
| *rh700_min* | Relative humidity at 700 hPa | relative_humidity @ 700 hPa | Daily minimum | % |
| *rh700_max* | Relative humidity at 700 hPa | relative_humidity @ 700 hPa | Daily maximum | % |
| *Lapse_mean* | Lower-tropospheric lapse rate ($T_{850} - T_{500}$) | temperature @ 850, 500 hPa | Daily mean | K |
| *Lapse_max* | Lower-tropospheric lapse rate ($T_{850} - T_{500}$) | temperature @ 850, 500 hPa | Daily maximum | K |
| *Thick_mean* | 1000–500 hPa geopotential thickness | geopotential @ 1000, 500 hPa | Daily mean | m |
| *w700_mean* | Vertical velocity ($\omega$) at 700 hPa | vertical_velocity @ 700 hPa | Daily mean | Pa $s^{-1}$ |
| *w500_min* | Vertical velocity ($\omega$) at 500 hPa | vertical_velocity @ 500 hPa | Daily minimum | Pa $s^{-1}$ |
| *wspd700_mean* | Wind speed at 700 hPa | u/v_component_of_wind @ 700 hPa | Daily mean | m $s^{-1}$ |
| *wspd850_mean* | Wind speed at 850 hPa | u/v_component_of_wind @ 850 hPa | Daily mean | m $s^{-1}$ |
| *wspd925_mean* | Wind speed at 925 hPa | u/v_component_of_wind @ 925 hPa | Daily mean | m $s^{-1}$ |
| *dir700_mean* | Wind direction at 700 hPa | u/v_component_of_wind @ 700 hPa | Daily mean | deg |
| *dir850_mean* | Wind direction at 850 hPa | u/v_component_of_wind @ 850 hPa | Daily mean | deg |
| *dir925_mean* | Wind direction at 925 hPa | u/v_component_of_wind @ 925 hPa | Daily mean | deg |
| *shear_mean* | Bulk vertical wind shear ($V_{500} - V_{850}$) | u/v_component_of_wind @ 850, 500 hPa | Daily mean | m $s^{-1}$ |

**Table 1**. Meteorological predictor variables (27) derived from the ECMWF reanalysis. Feature names correspond to the variable names used in the model; the ERA5 source column gives the reanalysis field(s) from which each feature is computed. Pressure-level fields are given at their respective levels (hPa). Six additional fields present in the daily feature files (vpd_max, wind90, t2m_max, rh_min, dpd_max, dpd_mean) were excluded from the final configuration on the basis of permutation-importance analysis and are not listed.

### *2.3. Land-Surface and Fire-History Predictors*

Static and slowly varying land-surface predictors describe fuel type and condition. Land cover is provided by the annual MODIS MCD12C1 IGBP classification (Friedl and Sulla-Menashe 2022) and assigned to the 0.1° grid by nearest-neighbor regridding to preserve categorical classes. Land cover is supplied as a single channel of IGBP class indices, despite the implied numerical ordering of categorical classes. A sensitivity test using one-hot encoding, together with improved handling of missing values, produced only small changes of mixed sign: detection skill was essentially unchanged (CSI within ±0.005), while intensity and MSE skill scores improved slightly in some cases and declined in others. The main results are therefore insensitive to the encoding choice, although one-hot encoding will be adopted in future versions.

Fire history enters through two complementary predictors. The first is the observed daily mean FRP field (MeanFRP_t) from GBBEPx v4, supplied for each of the 21 days up to and including initialization, providing both the current fire state and its recent evolution. The second is a duration-of-fire predictor (dof), defined as the number of consecutive days with active fire (FRP > 0) at each pixel and reset to zero when fire ceases. This distinguishes newly ignited pixels from persistent, mature fires.

Per-pixel day-of-year FRP climatological statistics provide an additional prior on local fire behavior. For each pixel and calendar day, the median and 90th percentile of FRP are calculated from fire days (FRP ≥ 0.5 MW) in the 2012–2023 GBBEPx record, requiring at least three qualifying observations; otherwise the value is set to zero. These fields (frp_clim_median and frp_clim_p90) represent the characteristic seasonal intensity distribution at each pixel. The 90th-percentile field also serves as the reference for the climatology-constrained intensity cap described in Section 3.4, so the same climatology informs both model input and output constraint.

| Feature name | Description | Source | Type | Units |
|---|---|---|---|---|
| *igbp* | IGBP land-cover classification | MODIS MCD12C1 (annual) | Static (annual) | class |
| *vhi* | Vegetation Health Index | Weekly VHI product | Dynamic (weekly) | dimensionless |
| *dof* | Duration of fire: consecutive days of active fire (FRP > 0) at the pixel, reset to zero when fire ceases | Derived from GBBEPx v4 | Dynamic | days |
| *MeanFRP_t* | Observed daily mean FRP at initialization | GBBEPx v4 | Dynamic | MW |
| *frp_clim_median* | Per-pixel day-of-year median FRP on fire days | GBBEPx (2012–2023) | Static climatology | MW |
| *frp_clim_p90* | Per-pixel day-of-year 90th-percentile FRP on fire days | GBBEPx (2012–2023) | Static climatology | MW |

**Table 2**. Land-surface and fire-history predictor variables (6). The IGBP land-cover classification is a static annual field; the Vegetation Health Index is a weekly product held constant across the days of each ISO week; the duration-of-fire and climatology features are derived from the GBBEPx FRP record.

The three FRP-based predictors (antecedent FRP and the two climatological fields) are log-transformed on input to accommodate their heavy-tailed distribution. During training, the antecedent FRP field is randomly withheld on a small fraction of samples so that the model does not become overly reliant on persistence and retains skill when recent fire observations are unavailable. The complete set of land-surface and fire-history predictors is listed in Table 2.

### *2.4. Training and Evaluation Periods*

The model is trained on a continuous three-year record from 1 January 2020 to 31 December 2022. Accounting for the 21-day input history and the seven-day forecast horizon, this yields 1069 initialization days spanning 21 January 2020 to 24 December 2022, covering multiple complete fire seasons in both hemispheres. Each domain model is trained for 200 epochs. A 21-day window is used because fire seasons last several months and individual fires can burn for more than a month, so a longer daily history is needed to capture their evolution. Training uses 2020–2022 rather than the full GBBEPx record because of computational constraints and because fire regimes have changed rapidly, with extreme fires becoming more frequent and intense (Iglesias et al. 2022; Cunningham et al. 2024).

Evaluation is performed over two complete calendar years, 2023 and 2024, both strictly independent of the training period. A seven-day forecast is generated every day from the input fields listed in Tables 1 and 2, and each is verified at all lead times from one to seven days against the corresponding GBBEPx v4 observations. This yields a temporally dense verification sample spanning two complete annual cycles and supports the lead-time-resolved and month-resolved assessments of Section 5. The 2024 year additionally lies outside the 2012–2023 climatological record used for the climatological predictors and the intensity constraint.

## 3. Methods

### *3.1. Model Overview*

The prediction model is a spatiotemporal graph neural network adapted from the GraphCast framework (Lam et al. 2023), referred to here as FRP-GraphCast-Lite. At forecast initialization, the model ingests a 21-day history of the predictor fields over a region and produces, for each land grid cell, a prediction of FRP at each of the seven subsequent daily lead times. Inference is performed on overlapping 128 × 128-cell spatial patches (stride 96 cells in each direction), which are reassembled into the full regional field by averaging over the overlap regions; this tiling keeps the graph at a tractable size while providing halo overlap so that patch boundaries are not visible in the reassembled field.

The model architecture comprises three components. First, an input projection maps the per-cell predictor vector to a 256-dimensional latent representation. Second, two residual graph-convolution layers propagate information among neighboring cells; the land surface is represented as a graph in which nodes correspond to grid cells and each interior node is connected to its four nearest neighbors (north, south, east, west), so that message passing captures the local spatial structure of fire spread and the organization of fire activity along terrain and vegetation gradients. Unlike the multi-mesh icosahedral graph used in GraphCast, the present model constructs a regular four-connected lattice graph on the native latitude–longitude grid within each spatial patch. This choice reflects the local, short-range nature of fire spread, which does not require the long-range spherical connectivity needed for atmospheric wave propagation, and keeps the graph tractable at 0.1° resolution. The graph formulation also allows the model to be transferred directly to the UFS Finite-Volume Cubed-Sphere (FV3) cubed-sphere grid and the Model for Prediction Across Scales (MPAS) unstructured mesh. Each graph-convolution layer combines a self-transformation, a degree-normalized neighbor aggregation, and a residual skip connection, followed by a GELU nonlinearity. Third, a temporal encoder integrates the multi-day history: the per-day latent fields are stacked along the time axis and processed by a two-layer one-dimensional convolutional network (kernel size 3, GELU activations), and the final time step of its output is taken as the integrated latent state. Two linear prediction heads then map this state to the seven-day forecast, one producing fire-occurrence logits and the other producing fire intensity, each of shape (7 lead days × grid cells).

To address the extreme sparsity and heavy-tailed distribution of FRP, the prediction head adopts a hurdle (two-part) formulation. The classification head produces the probability of fire occurrence in each cell, and the regression head produces the conditional fire intensity, parameterized in log space: the predicted intensity is expm1(μ), where μ is the regression output constrained to a physically bounded range (Section 3.4). The final predicted FRP is the product of the predicted occurrence probability and the predicted intensity, after which predictions below a fixed gate of 0.15 MW are set to zero to suppress low-confidence detections. This separation prevents the regression component from being overwhelmed by the predominance of zero values and allows the occurrence and intensity sub-problems to be optimized with objectives appropriate to each (Section 3.2).

*3.2. Loss Function*

Model parameters are optimized with a composite objective designed for the statistical structure of FRP. The total loss is a weighted sum of five terms:

$$\mathbf{L} = 2.0 \cdot L_{\mathrm{cls}} + 10.0 \cdot L_{\mathrm{reg}} + 50.0 \cdot L_{\mathrm{bg}} + 1.0 \cdot L_{\mathrm{vol}} + 0.5 \cdot L_{\mathrm{spec}} \qquad (1)$$

where $L_{\mathrm{cls}}$, $L_{\mathrm{reg}}$, $L_{\mathrm{bg}}$, $L_{\mathrm{vol}}$, and $L_{\mathrm{spec}}$ are the occurrence, intensity, background, volume, and spectral terms defined below, with weights determined empirically during development and held fixed thereafter.

Let $y$ denote the observed FRP at a grid cell (MW), $z$ the classification logit, and $\mu$ the regression output. The predicted FRP is formed as

$$\hat{\mathbf{y}} = \sigma(z) \cdot \mathrm{expm1}(\mathrm{clamp}(\mu, -10, \mu_{\mathrm{max}})) \qquad (2)$$

where σ is the logistic function and $\mathrm{expm1}(x) = e^x - 1$ is the inverse of $\mathrm{log1p}(x) = \log(1 + x)$. This transform pair is used because FRP spans several orders of magnitude and contains many exact zeros, for which $\mathrm{log1p}(0) = 0$ is well defined. The regression head therefore predicts intensity in log space and equation (2) maps it back to physical units; the network itself is nonlinear throughout, and $\mu_{\mathrm{max}}$ is the per-region bound of Section 3.4.

**Occurrence.** $L_{\mathrm{cls}}$ is a focal loss applied to $z$ against the binary target $1[y > 0]$:

$$L_{\mathrm{cls}} = \mathrm{mean}(\ \alpha(1 - p_t)^{\gamma} \cdot \mathrm{BCE}(z, 1[y > 0])\ ), \quad p_t = \exp(-\mathrm{BCE}) \qquad (3)$$

with $\alpha = 0.25$ and $\gamma = 2$, which down-weights the abundant, easily classified background cells and concentrates learning on fire pixels.

**Intensity.** The regression term is evaluated in physical units (MW). A per-pixel smooth-L1 (Huber) error is computed first and then scaled by an intensity weight and an asymmetry factor:

$$h = \mathrm{SmoothL1}(\hat{\mathbf{y}}, y; \beta = 1) \qquad (4)$$

$$r = h \cdot (\log(1 + y) + 1)^2 \cdot a(d), \quad d = \hat{\mathbf{y}} - y, \quad a(d) = 10 \text{ if } d < 0, \text{ else } 1 \qquad (5)$$

$$L_{\mathrm{reg}} = \text{mean of the largest 5\% of } r \qquad (6)$$

The intensity weight rises from 1.0 at $y = 0$ to 24.3 at $y = 50$ MW and 90.6 at $y = 5000$ MW, so a large fire contributes roughly ninety times the weight of a background cell. The asymmetry factor penalizes under-prediction ten times more strongly than over-prediction, reflecting the operational cost of missing intense, smoke-producing events. The top-k selection concentrates optimization on the most consequential errors rather than diluting it across the vast majority of near-zero cells.

**Background.** $L_{bg} = \text{mean}(\hat{\mathbf{y}} \cdot 1[y = 0])$ penalizes any nonzero prediction over cells with no observed fire.

**Volume.** With $\hat{V}$ and $V$ the predicted and observed FRP summed over each patch, and $D = \hat{V} - V$,

$$L_{vol} = \text{mean}(|D| \cdot a_v(D)), \quad a_v(D) = 5 \text{ if } D < 0, \text{ else } 1 \qquad (7)$$

**Spectral.** $L_{spec}$ compares the two-dimensional spatial spectra of the predicted and observed fields, as the mean squared difference of the log-transformed magnitudes of their orthonormal real-input FFTs:

$$L_{spec} = \text{mean}(\ [\log(1 + |F(\hat{\mathbf{y}})|) - \log(1 + |F(y)|)]^2\ ) \qquad (8)$$

encouraging agreement in spatial structure independent of exact pixel placement.

| Symbol | Meaning | Units | Value |
|---|---|---|---|
| y | Observed FRP at a grid cell | MW | — |
| ŷ | Predicted FRP, eq. (2) | MW | — |
| z, μ | Classification logit; regression output | — | — |
| μmin, μmax | Regression clamp bounds (Sec. 3.4) | log MW | −10.0; 10.0 (SEAsia, EMEA), 11.0 (others) |
| α, γ | Focal loss parameters | — | 0.25, 2.0 |
| $(\log(1+y)+1)^2$ | Intensity weight | — | 1.0 → 90.6 for y = 0 → 5000 MW |
| a(d) | Regression asymmetry factor | — | 10.0 under-prediction; 1.0 otherwise |
| av(D) | Volume asymmetry factor | — | 5.0 under-prediction; 1.0 otherwise |
| k | Top-k fraction, eq. (6) | — | 5% (minimum 10 pixels) |
| — | Fire threshold for classification and background target | MW | y > 0 |
| F | Orthonormal 2-D real FFT | — | — |
| — | Term weights in eq. (1) | — | 2.0, 10.0, 50.0, 1.0, 0.5 |

**Table 3.** Loss components and parameters

*3.3. Region-based Training Strategy*

Global fire regimes are sufficiently distinct over regions — in seasonality, fuel type, and characteristic fire intensity — that a single global model cannot represent them simultaneously without the most extreme regimes destabilizing optimization. The land surface is therefore partitioned into five fire-regime domains: North America (12°–85° N, 170°–50° W), South America (60° S–12° N, 90°–30° W), Europe–Middle East–Africa (EMEA; 40° S–85° N, 30° W–60° E), Southeast Asia (10° S–85° N, 60°–180° E), and the Australia–Pacific sector (55°–10° S, 60°–180° E). A separate model is trained for each domain. The five share an identical architecture, predictor set, and loss formulation, but their parameters are learned independently; no weights are shared across domains. Global fields are produced at inference by mosaicking the five regional predictions, with consistent longitude-wrapping handling so that the eastern-hemisphere domains align correctly on a common grid; where domains meet at their boundaries, the maximum of the overlapping predictions is taken.

A small number of settings are tuned per domain to reflect the differing intensity distributions of, for example, boreal crown fires and tropical peatland burning; all others are common. Southeast Asia, dominated by intense equatorial peatland and agricultural burning, uses the tightest clip on FRP inputs and targets (1500 MW), with an intermediate value in the Australia–Pacific domain (3000 MW) and 5000 MW elsewhere. Southeast Asia and EMEA additionally use a tighter upper bound on the predicted intensity mean during training, and a tighter multiplier on the climatological intensity cap at inference (3× and 4× the historical per-pixel 90th percentile, respectively, versus 5× elsewhere). Southeast Asia additionally trains at a reduced initial learning rate ($10^{-7}$ versus $5 \times 10^{-6}$ in all other domains), which was necessary for stable convergence in that domain. All models otherwise share a common configuration: a 21-day input history, 200 training epochs, the AdamW optimizer with weight decay of $10^{-4}$ and a cosine-annealed learning-rate schedule, and per-epoch sampling of fire-containing patches drawn in proportion to an intensity-based weight so that active-fire days are emphasized. During training, the antecedent observed-FRP and duration-of-fire inputs are jointly zeroed on 5% of samples so that the model does not become overly reliant on persistence of recent fire activity.

### *3.4. Climatology-Constrained Intensity Cap*

A purely data-driven model has no built-in physical constraint preventing implausible intensity excursions when extrapolating to rare combinations of predictors. This is particularly relevant for FRP, whose distribution is heavy-tailed and dominated by a small number of extreme events, and whose intensity-weighted, under-prediction-penalizing loss (Section 3.2) can favor inflated predictions on outlier days. To enforce physical plausibility, the model applies a climatology-constrained per-pixel intensity cap at inference time.

The predicted FRP in each cell is limited to a region-specific multiple of that cell's historical day-of-year 90th-percentile FRP climatology, subject to an absolute floor:

per-pixel cap = max(frp_clim_p90(DOY) × CLIM_CAP_FACTOR, CLIM_CAP_FLOOR),
prediction = min(prediction, per-pixel cap).

The cap uses the valid-time day of year for each forecast lead. The factor is set to 5 for North America, South America, and Australia–Pacific, where genuine extremes rarely exceed five times the historical 90th percentile, and to 4 for EMEA and 3 for Southeast Asia, whose climatological upper tails are elevated by extreme burning during the training period and would otherwise permit near-record predictions on ordinary days. The climatology is calculated from the 2012–2023 GBBEPx record. Because it serves both as a predictor and as the output constraint and includes one evaluation year (2023), this overlap is noted as a limitation.

The floor is set to 5000 MW, matching the upper clip applied to FRP targets during training, so the constraint cannot bind below the supervised range; it also governs where the climatology is undefined. An earlier 200 MW floor acted as a near-global ceiling for this reason and was abandoned. One isolated pixel (67.5°N, 179.8°E), with a single fire day in 4627 observations, produced persistently high predictions without observational support and is masked in both prediction and verification.

The same climatological sparsity produced one isolated pathological pixel at 67.5° N, 179.8° E, where persistently high FRP was predicted despite zero observed fire during evaluation. The pixel recorded only one fire day in 4627 valid observations during 2012–2023—enough to pass the persistent-zero mask but insufficient to define a day-of-year climatology—leaving the 5000 MW floor as the only intensity constraint despite essentially no observational support. This pixel is therefore masked to zero in both prediction and verification. With the floor at 5000 MW, the cap rarely engages: outside Southeast Asia it was active on 1.6% of region-days, removing at most 6.8% of predicted radiative power, and in Southeast Asia it acted almost entirely on persistently non-burning cells removed by the persistent-zero mask.

Two further safeguards are applied: μ is clamped before the expm1 transform (equivalent to about 60,000 MW, or 22,000 MW for EMEA and Southeast Asia), and a final absolute cap of 20,000 MW is imposed. A regional multiplicative bias scale is 1.0 everywhere except EMEA (0.8), where it reduces a systematic positive intensity bias without retraining.

### *3.5. Operational Configuration and Inference*

Global fields are produced by applying each regional model to the tiled patches of Section 3.1 and merging the regions as described in Section 3.3. Post-processing is applied in a fixed order: the 0.15 MW gate, the climatology cap, the regional bias scale, and the 20,000 MW output cap; persistently non-burning cells are masked to zero.

In operational use the system is intended to be driven by forecast meteorology, enabling genuine multi-day predictive smoke forcing rather than persistence. The verification reported here uses reanalysis meteorology as predictor input in order to isolate the skill of the FRP model from errors in the meteorological forecast, and therefore represents an upper bound on the skill achievable with forecast meteorology of comparable quality; the magnitude of that reduction has not yet been quantified and will be assessed by driving the model with GFS/GEFS forecast meteorology (see the final paragraph of Section 6).

## 4. Verification Framework

Predictions are verified against the GBBEPx observed FRP field over the independent evaluation period using a dual-resolution strategy. Verification is performed at the native 0.1° resolution, which emphasizes exact spatial placement, and at 1° resolution after spatial aggregation, which is more representative of the scale relevant to regional aerosol and weather prediction and less sensitive to small displacement errors. Skill is evaluated separately at lead times of one to seven days and also pooled across leads for an overall summary.

Two complementary classes of metrics are used. Categorical detection skill treats fire occurrence as binary above a threshold of 0.5 MW per 0.1° pixel, consistent with the threshold used in constructing the GBBEPx climatology, and is summarized by the probability of detection (POD), false-alarm ratio (FAR), critical success index (CSI), Heidke skill score (HSS), equitable threat score (ETS), and frequency-bias index (FBI). Continuous intensity skill over fire-affected cells is evaluated using root-mean-square error (RMSE), mean absolute error (MAE), mean bias error (MBE; predicted minus observed), Pearson correlation, Kling–Gupta efficiency (KGE), and Nash–Sutcliffe efficiency (NSE). All metrics are computed over grid cells that record fire at least once in the climatological record, excluding persistently non-burning cells. Continuous metrics are weighted by cosine of latitude, so each cell contributes in proportion to its area, while categorical scores use unweighted contingency counts.

Forecast skill is also assessed relative to persistence of the initialization-time observed FRP held fixed across the forecast period. This reproduces the operational assumption that a single recently observed emission field drives the entire forecast and therefore provides the most relevant reference for assessing forecast value. Skill relative to persistence is expressed as 1 − MSE_model / MSE_persistence, for which positive values indicate lower error than the persisted field.

Verification and any subsequent conversion to emissions consider only values above 0.5 MW per 0.1° pixel, consistent with the fire threshold used in constructing the FRP climatology in this study. This has negligible effect on observations: in 2023, values between 0 and 0.5 MW represented only 214 of 3.62 million nonzero grid cells (0.006%).

| **Metric** | **Type** | **Quantity assessed** | **Range** | **Better** |
|---|---|---|---|---|
| POD | Categorical | Fraction of observed fires detected | 0 to 1 | Higher |
| FAR | Categorical | Fraction of predicted fires that are false | 0 to 1 | Lower |
| CSI | Categorical | Overall detection skill (threat score) | 0 to 1 | Higher |
| HSS | Categorical | Skill relative to random chance | ≤ 1 | Higher |
| ETS | Categorical | Chance-corrected threat score | −1/3 to 1 | Higher |
| FBI | Categorical | Predicted vs observed fire frequency | ≥ 0 | ≈ 1.0 |
| RMSE | Continuous | Magnitude of intensity error | ≥ 0 | Lower |
| MAE | Continuous | Mean absolute intensity error | ≥ 0 | Lower |
| MBE | Continuous | Systematic intensity bias | unbounded | ≈ 0.0 |
| Pearson r | Continuous | Linear association of intensity | −1 to 1 | Higher |
| KGE | Continuous | Combined correlation, bias and variability | ≤ 1 | Higher |
| NSE | Continuous | Variance explained relative to the observed mean | ≤ 1 | Higher |

**Table 3**. Verification metrics used in this study. Categorical metrics are computed on binary fire/no-fire fields above a 0.5 MW threshold; continuous metrics are computed over fire-affected cells. The Range column gives the attainable values of each score.

## 5. Results and Discussion

### *5.1. Overall Skill, Seasonal Cycle and Spatial Structure*

Figure 1 compares predicted and observed daily global-mean FRP at lead day +1, lead day +7, and averaged over all leads verifying on each date. At D+1 the model reproduces the seasonal cycle in both years: elevated activity from February through April, a minimum in May and June, the boreal-summer rise, an annual maximum in late August, and the autumn decline. Day-to-day fluctuations are followed rather than smoothed. Including all days, the median absolute daily bias at 0.1° is 0.048 MW per 0.1° pixel in 2023 and 0.044 in 2024, with 83% and 77% of days within ±0.1 MW; integrated over the year, predicted total radiative power is within 2% of observed in 2023 and 12% above it in 2024. The largest errors occur at the sharpest peaks and follow a consistent pattern: predicted peaks follow observed peaks by one to two days and exceed them. In 2023 the observed maximum of 0.76 MW on 23 August is followed by a predicted maximum of 0.80 MW on 25

August. In 2024 the forecast exceeds the observations throughout late August and early September, and each observed local peak is followed a day later by a larger predicted one (0.85 MW on 3 September against 0.95 MW on 4 September; 0.81 against 0.99 MW on 7–8 September), culminating in the predicted annual maximum of 1.01 MW on 10 September; the observed annual maximum, 0.88 MW, occurred earlier, on 21 August.

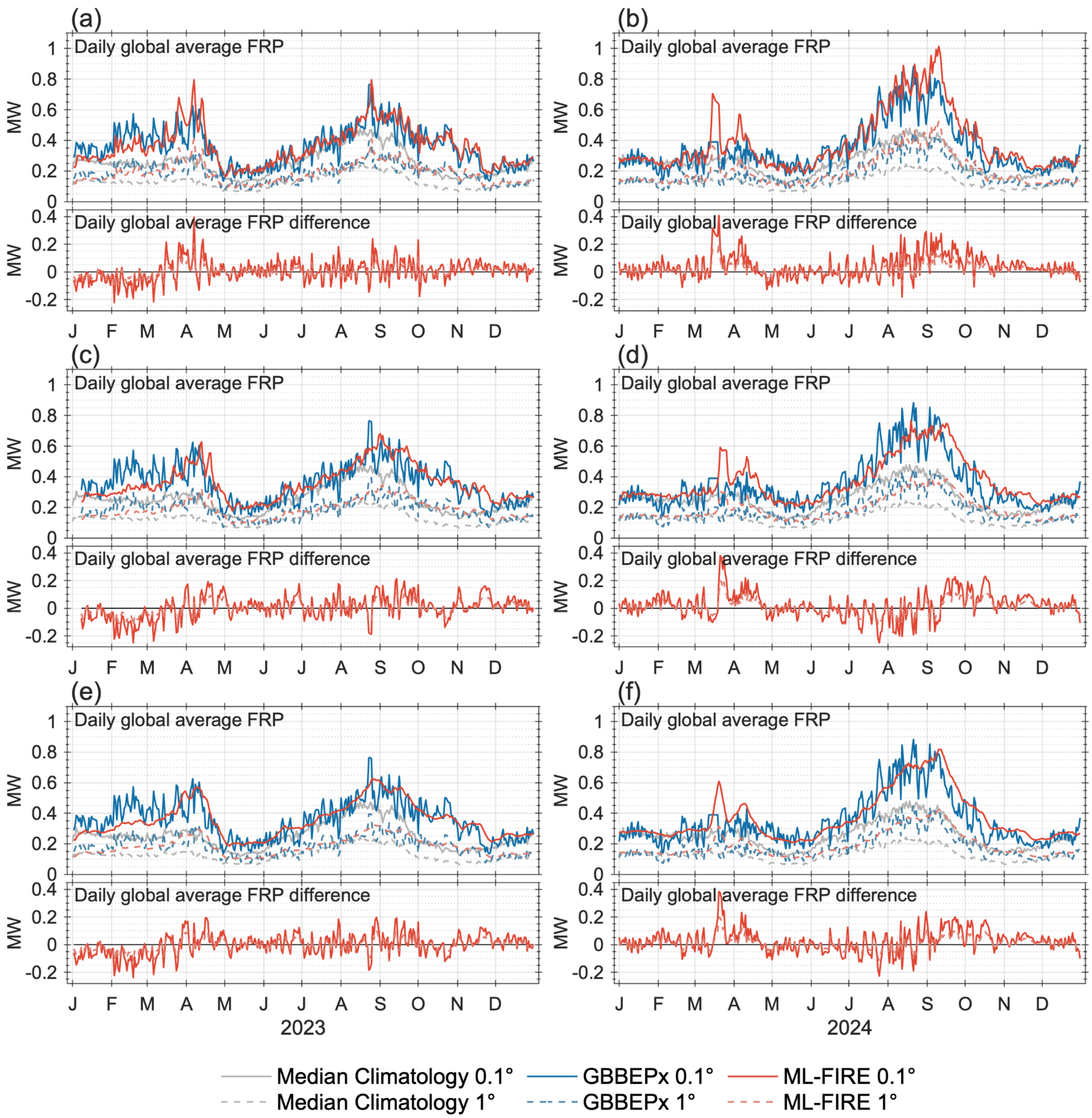

**Figure 1.** Daily global-mean fire radiative power (FRP; MW per 0.1° pixel) in 2023 (left column) and 2024 (right column). Rows show forecasts at lead day +1 (a, b), lead day +7 (c, d), and the mean of all forecasts (leads 1–7) verifying on each date (e, f). In each upper subpanel, ML-FIRE predictions (red) and GBBEPx observations (blue) are shown at the native 0.1° resolution (solid) and aggregated to 1° (dashed); grey lines show the GBBEPx median FRP climatology for the same day of year (2012–2023). The lower subpanels show the daily bias (predicted − observed) at 0.1°. Observed values are constant from 12 to 19 March 2024 because of an interruption in the operational GBBEPx data stream, not actual fire activity.

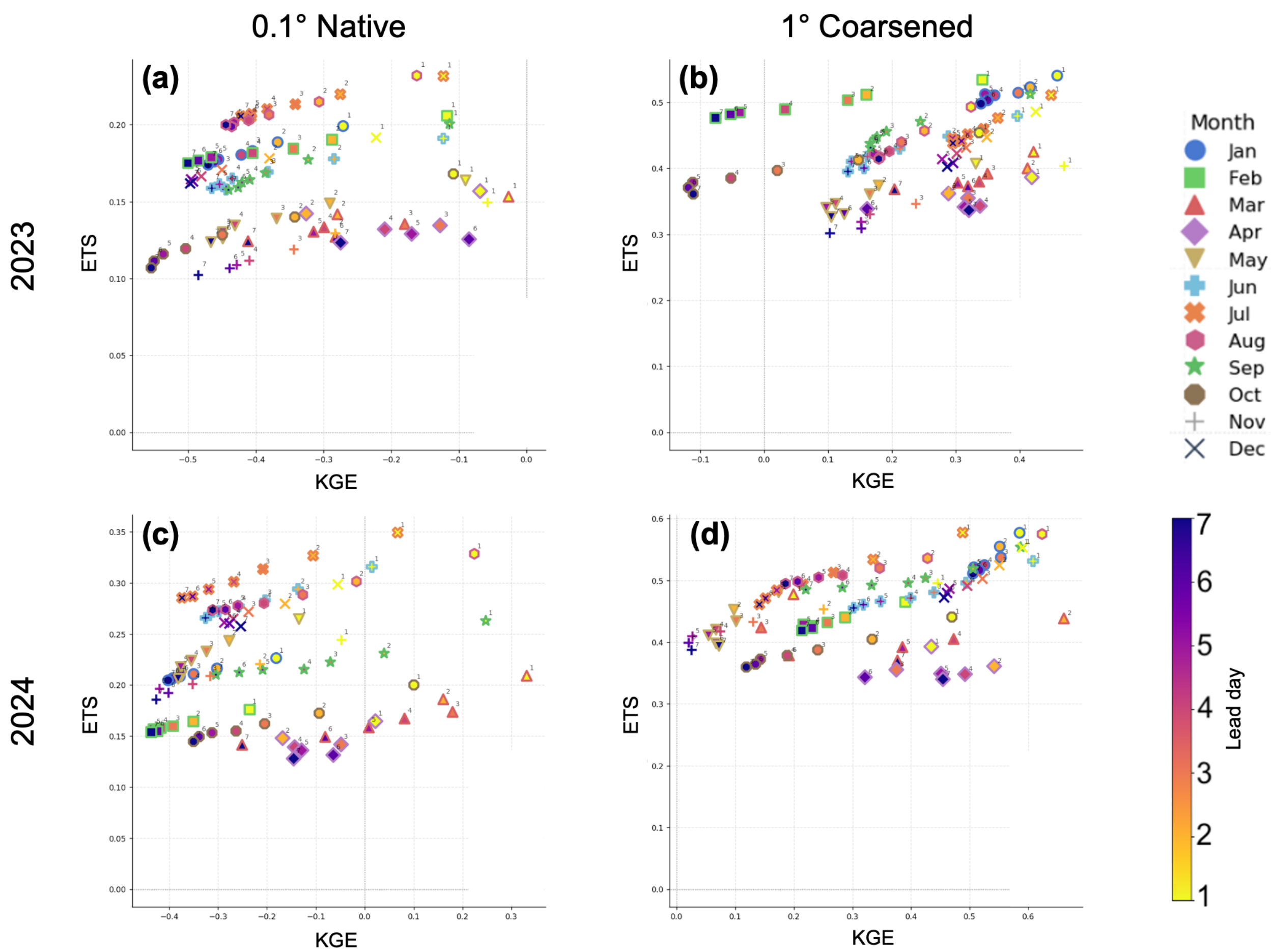

**Figure 2.** ETS versus KGE, one symbol per (month, lead day) pair. (a, b) 2023 at 0.1° and 1°; (c, d) 2024. Marker shape encodes month; color and the adjacent numeral both encode lead day (light yellow, 1 = D+1; dark purple, 7 = D+7). Axis ranges differ between panels.

The forecasts also capture year-to-year differences that the median climatology cannot: the stronger February–April activity of 2023 and the much larger August–September peak of 2024 both appear in the predictions, whereas the climatology is identical in both years and lies below the observations for most of each year. At D+7 and in the all-lead mean the seasonal cycle is retained, but the sharpest daily peaks are damped. The model therefore reproduces the timing and interannual variation of the fire season to within a few days, while lagging and amplifying the sharpest daily peaks at short lead. One feature of the 2024 panels requires comment. Between 12 and 19 March 2024 the observed series is constant (0.39 MW) because the GBBEPx files repeat an identical field during an interruption of the operational data stream. Excluding these days changes the 2024 statistics only marginally (median absolute bias 0.043 MW, 78% of days within ±0.1 MW, annual overestimate 11%).

Figure 2 summarizes joint detection and intensity skill as ETS against KGE, one symbol per (month, lead day) pair. At 1° the two years occupy overlapping regions: ETS spans 0.30–0.55 in 2023 and 0.34–0.59 in 2024, KGE spans −0.12 to 0.46 and 0.00 to 0.62, positive in 77 of 84 month–lead combinations in 2023 and in all 84 in 2024. The points follow a coherent diagonal along which detection and intensity skill increase together. At 0.1° the structure differs: ETS spans 0.10–0.23 and 0.13–0.35, KGE is negative in all 84 combinations in 2023 and in 72 of 84 in 2024, and the points separate into ETS bands rather than following a single diagonal. Detection and intensity skill are therefore less tightly coupled at the native resolution than after aggregation.

Figure 3 shows monthly metrics at one-day lead. Absolute errors scale with monthly fire activity: MAE at 0.1° peaks in April at 34.9 MW in 2023 and 22.8 MW in 2024, and is systematically smaller in 2024 across almost every month. Mean bias error at 0.1° is negative in every month of both years, from −11.3 to −16.4 MW in 2023 and from −3.4 to −11.1 MW in 2024, while at 1° monthly biases are within ±1 MW. Categorical skill shows a seasonal structure consistent between years: POD at 1° spans 0.56–0.77 and 0.57–0.81, FAR spans 0.26–0.46 and 0.26–0.43, and CSI spans 0.40–0.55 and 0.41–0.60, with the strongest detection from July to September. At 0.1° the scores are uniformly weaker (CSI 0.16–0.25 and 0.17–0.37). Notably, POD is comparable at the two resolutions while FAR is two to three times as high at 0.1°: the gap in CSI is driven by false alarms arising from imprecise placement rather than by missed fires. This scale dependence is characteristic of machine-learning forecast systems more generally. Deterministic models trained under mean-targeted objectives tend to reproduce the large-scale organization of a field more faithfully than its fine-scale structure, so that verification scores improve markedly under spatial aggregation; Selz et al. (2026) show that the effective resolution of current AI weather prediction models is several times coarser than their nominal grid spacing for the same reason. The Fractions Skill Score analysis below quantifies that scale for the present model.

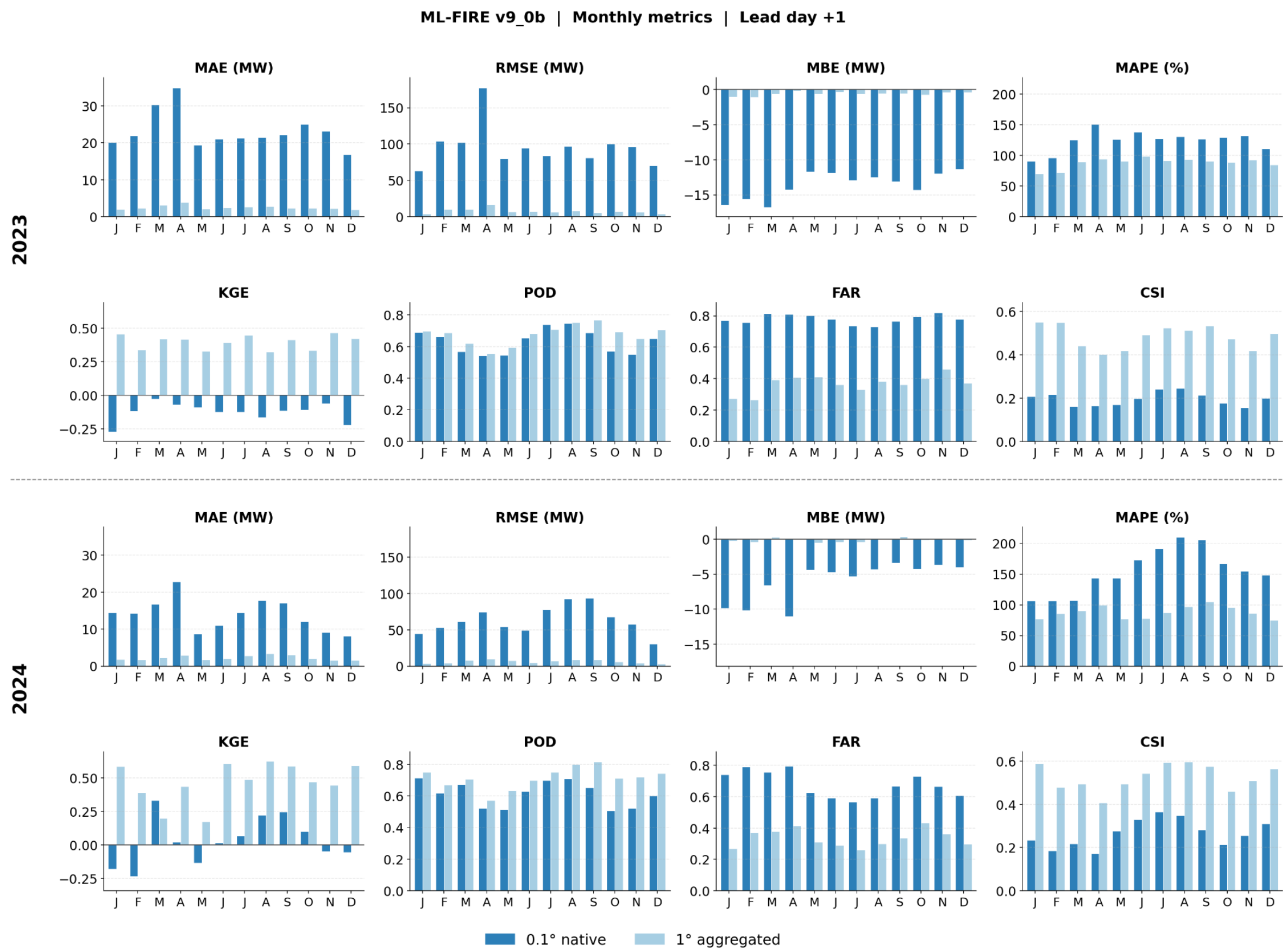

**Figure 3.** Monthly verification metrics at lead day +1 for 2023 (top) and 2024 (bottom), at 0.1° native (dark) and 1° aggregated (light) resolution. Continuous metrics are computed on cells with observed FRP above 0.5 MW per 0.1° pixel; categorical metrics use the same threshold to define fire.

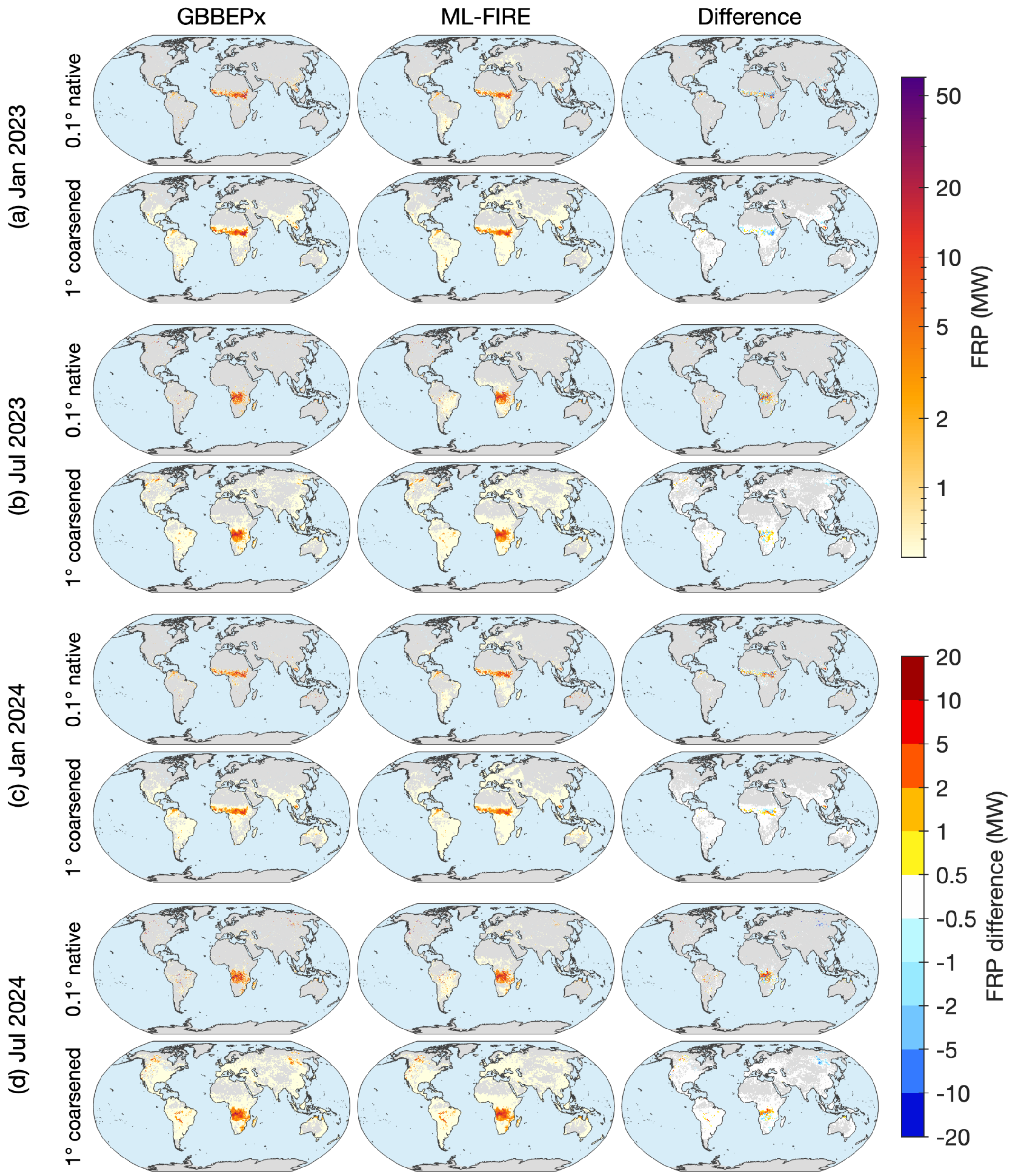

**Figure 4.** Monthly mean FRP at a one-day forecast lead time for (a) January 2023, (b) July 2023, (c) January 2024, and (d) July 2024. Columns show GBBEPx, ML-FIRE predictions, and their difference (ML-FIRE − GBBEPx). Within each panel, the upper and lower rows show the native 0.1° and coarsened 1° resolutions, respectively. Color scales denote FRP (top) and FRP differences (bottom).

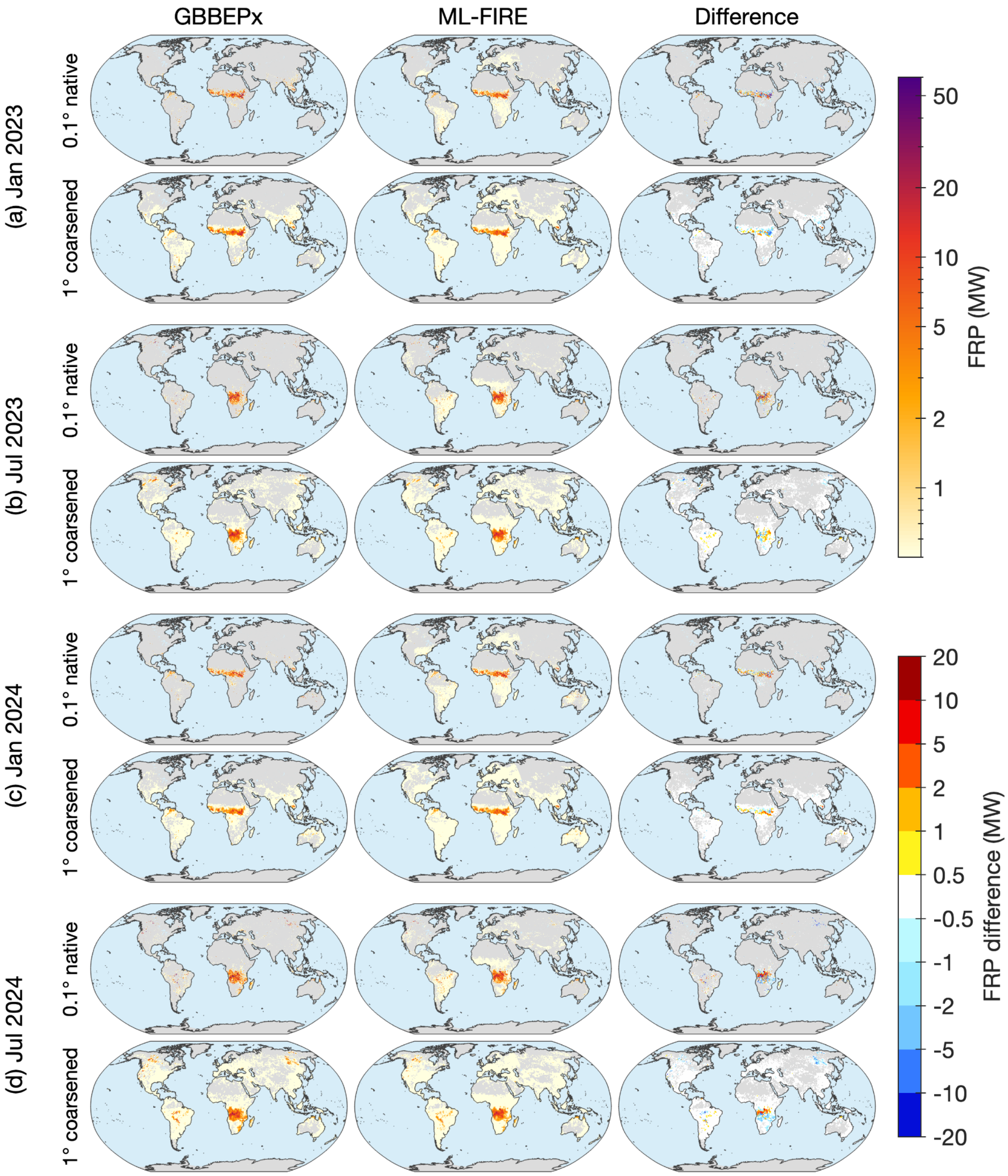

**Figure 5.** As in Figure 4, but for a seven-day forecast lead time.

Figures 4 and 5 show monthly mean fields for January and July of both years at one- and seven-day lead; April and October are shown in Figs. A1–A2. The model reproduces the location and seasonal migration of the major burning regions — the Sahel in January and central and southern Africa in July — and their relative intensity. The difference fields are close to zero over most of the domain but show a salt-and-pepper structure in the active regions at 0.1°, with adjacent positive and negative values of comparable magnitude. This is the spatial signature of displacement error:

fire is predicted in approximately the right place but not in exactly the right pixel. Aggregation to 1° removes most of this cancellation-prone structure, which is why the categorical scores improve so markedly with coarsening. One systematic omission is visible in the July panels of both years: scattered high-intensity pixels observed across eastern Siberia have no counterpart in the prediction, whereas comparable boreal fires over North America in the same month are captured. Since 2024 is the most distant year from the training period, this ordering cannot be explained by temporal proximity, and its cause has not been isolated.

### *5.2. Lead-Time Degradation: Detection versus Intensity*

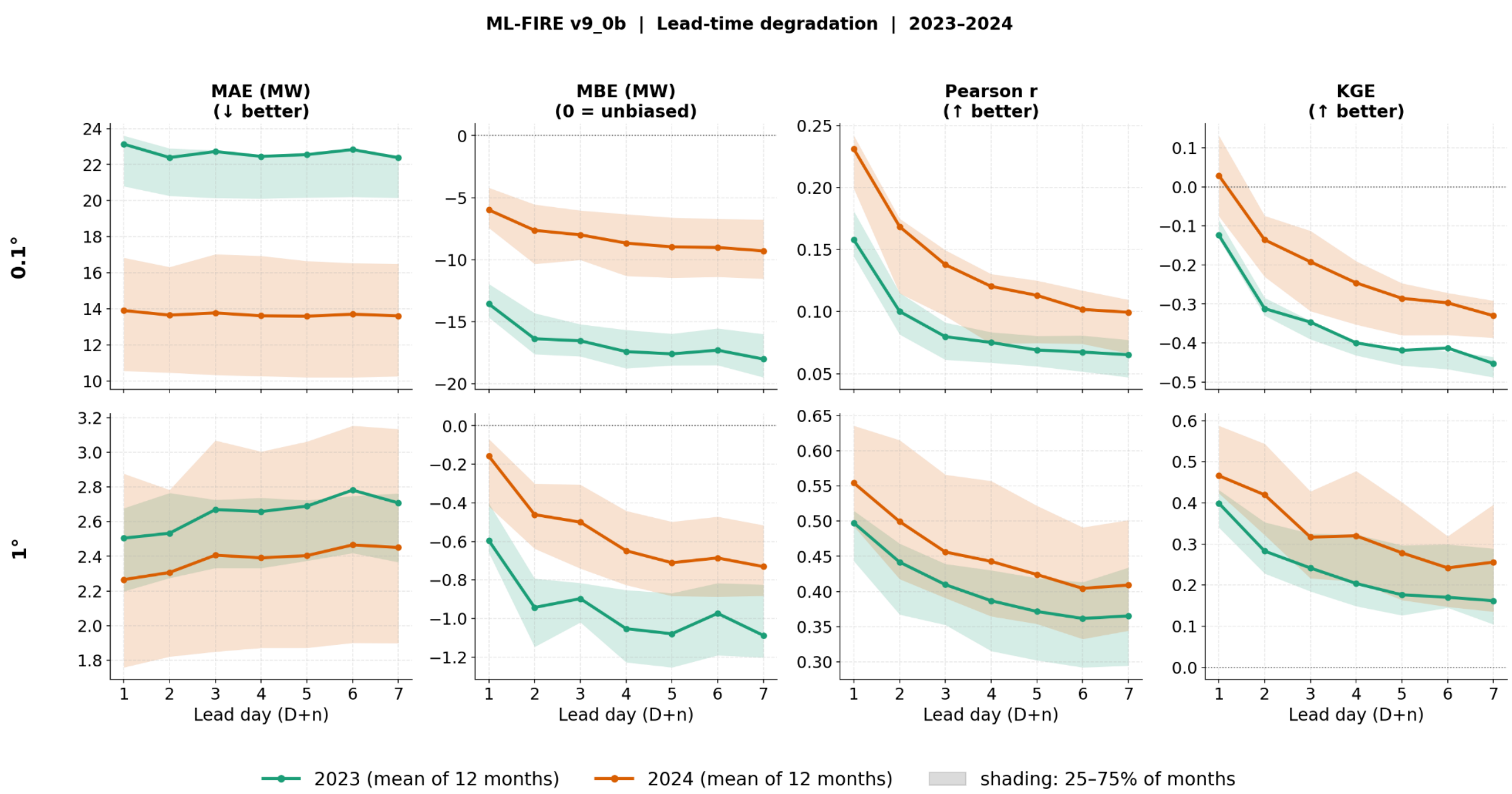

**Figure 6.** Lead-time degradation of verification metrics. Columns show mean absolute error (MAE), mean bias error (MBE), Pearson correlation (r) and Kling–Gupta efficiency (KGE); rows show 0.1° native (top) and 1° aggregated (bottom) resolution. Lines show the mean of the 12 monthly values for 2023 (green) and 2024 (orange); shading spans the 25th–75th percentiles across months. Metrics are computed on cells with observed FRP above 0.5 MW per 0.1° pixel at each resolution; MAE and MBE are in MW per 0.1° pixel. Vertical scales differ between panels.

Figure 6 shows how four verification metrics, computed on fire-affected cells, change across the seven-day forecast window in both years and at both resolutions; each line is the mean of the 12 monthly values, and the shading spans the interquartile range across months. At 0.1°, absolute errors are nearly flat with lead time: MAE stays near 23 MW per 0.1° pixel in 2023 and 14 MW in 2024; across individual months it ranges from 17 to 35 MW in 2023 and from 8 to 23 MW in 2024, changing by no more than about 1 MW between D+1 and D+7 within any single month. At 1° MAE rises modestly, by about 8% over the window. The typical per-pixel error magnitude is therefore set by the fire-size distribution rather than by lead-time drift. The bias, by contrast, is negative throughout and grows with lead time: at 0.1° MBE goes from about −14 to −18 MW in 2023 and from −6 to −9 MW in 2024, so the model increasingly underestimates the intensity of burning pixels at longer leads. At 1° Pearson r falls from 0.42–0.70 at D+1 to 0.19–0.59 at D+7 in 2023; at 0.1° it falls from 0.11–0.20 to 0.02–0.11.

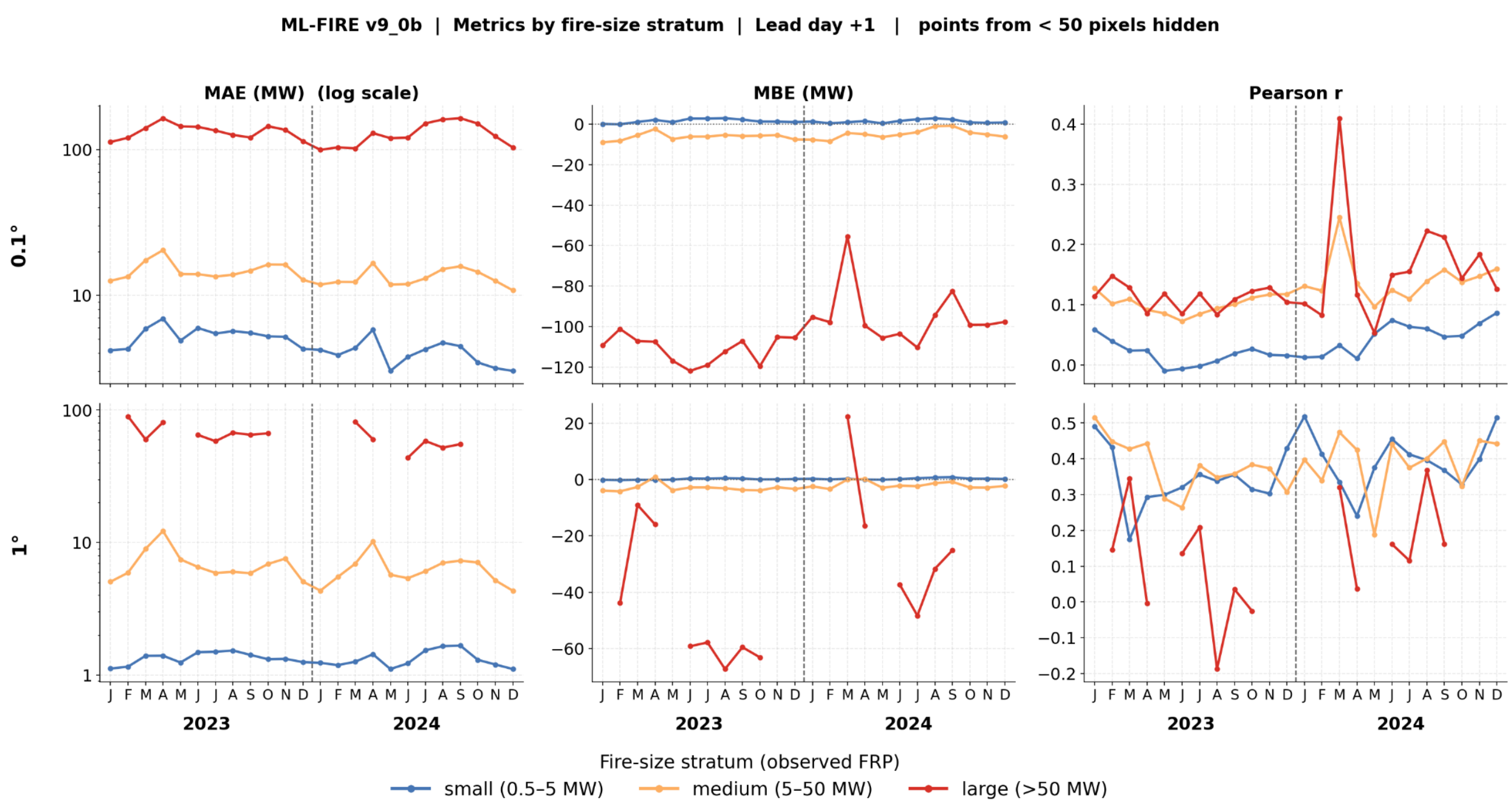

**Figure 7.** Verification metrics at one-day lead, stratified by observed fire size, for each month of 2023 and 2024 (separated by the dashed line). Columns show mean absolute error (MAE, log scale), mean bias error (MBE) and Pearson correlation (r); rows show 0.1° native (top) and 1° aggregated (bottom) resolution. Lines show the small (0.5–5 MW), medium (5–50 MW) and large (>50 MW) strata, defined on the observed FRP at each resolution; MAE and MBE are in MW per 0.1° pixel. Months in which a stratum contains fewer than 50 cells are omitted, which accounts for the gaps in the large stratum at 1°. Vertical scales differ between panels.

Taken with the categorical results, the asymmetry between the two classes of metric is the central finding of this evaluation. Between D+1 and D+7 at 0.1°, ETS declines by a median of only 0.032 in 2023 and 0.054 in 2024, while KGE declines by 0.33 and 0.36, an order-of-magnitude difference, in every month of both years without exception. The model retains its ability to identify which pixels are burning across the forecast window; what degrades is its ability to reproduce how strongly they burn. The contrast between flat MAE and declining correlation is a variance effect: the standard deviation of the predicted field falls from 6.7 to 4.0 MW between D+1 and D+7, against an observed 10.2 MW, as the forecast converges toward its conditional mean. Skill is consistently higher in 2024 than in 2023, with lower MAE, a smaller bias, and higher correlation and KGE at every lead. It is also substantially higher at 1° than at 0.1°: correlation at D+1 rises from 0.16–0.23 to 0.50–0.55, and KGE remains positive at all leads at 1° (0.16–0.47), whereas at 0.1° it is near or below zero, indicating that much of the 0.1° error reflects the placement of fire intensity among neighbouring pixels rather than its regional amount.

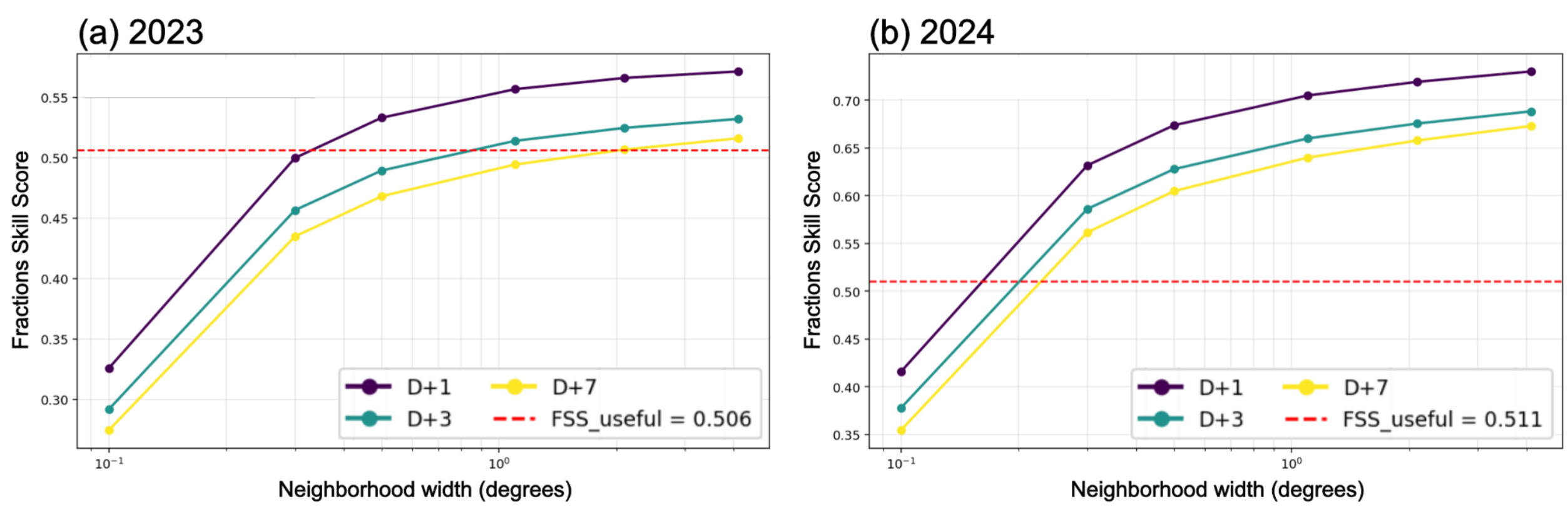

**Figure 8.** Fractions Skill Score as a function of neighborhood width, by forecast lead day. (a) 2023; (b) 2024. FSS evaluates the fraction of fire pixels within a square neighborhood rather than pixel-by-pixel agreement, so a forecast displaced by a small distance is not penalized twice — once as a miss and once as a false alarm — as it is by pointwise categorical scores. The dashed red line marks the useful-skill threshold $0.5 + 0.5 \cdot f$, where $f$ is the observed base rate of fire pixels; the neighborhood width at which a curve crosses it is the model's effective resolution at that lead. The horizontal axis is logarithmic.

Figure 7 stratifies the verification by observed fire size and approaches the same separation between detection and intensity from the other direction. Error magnitude rises steeply with fire size: at 0.1° in 2023, MAE is 4–7 MW for small fires, 13–20 MW for medium fires and 113–164 MW for large fires, with 2024 values of 3–6, 11–17 and 100–165 MW. The sign of the bias also changes with fire size. Small fires are slightly over-predicted, medium fires under-predicted by a few MW, and large fires strongly under-predicted in every month of both years, with MBE between −100 and −120 MW at 0.1° in 2023 and between −80 and −110 MW in 2024; the one exception, about −55 MW in March 2024, coincides with the interruption in the GBBEPx data stream described above. Averaged over the year, the large stratum is predicted at 0.24 of the observed intensity in 2023 (36 MW against 147) and 0.33 in 2024 (47 MW against 142). Correlation within each stratum is low at 0.1° (mostly below 0.2) but rises to 0.3–0.5 for small and medium fires at 1°; the large stratum at 1° contains too few cells in many months for a stable estimate.

Detection runs counter to this pattern. Probability of detection increases monotonically with observed intensity, from 0.56 for small fires through 0.67 for medium to 0.78 for large in 2023, and from 0.54 through 0.72 to 0.83 in 2024. The model locates large fires more reliably than small ones, and does so slightly better in 2024. The negative bias in the large stratum therefore reflects an error in predicted magnitude, not a failure to detect.

The climatology cap (Section 3.4) is not responsible for this deficit, since it rarely engages and removes far too little radiative power to account for an under-prediction of 67–76%. More plausible causes are the sparsity of extreme fires in the training data and the per-region clipping of FRP inputs and targets. A related diagnostic concerns the false alarms that dominate the 0.1° FAR. They are not low-intensity noise that a higher detection threshold would remove: predictions below 2 MW account for the majority of false-alarm pixels but only 8% of false-alarm radiative power, whereas pixels above 20 MW account for 43%. Removing half the false-alarm mass would require a threshold that also discards 90% of correctly located fires. The false alarms are fires of plausible intensity placed in the wrong location, a displacement error rather than a magnitude error, consistent with the salt-and-pepper structure of Figures 4 and 5 and with the Fractions Skill Score, which crosses the useful-skill threshold at a neighborhood width of only two to three grid cells.

The scale at which displacement ceases to matter can be quantified directly. Figure 8 shows the Fractions Skill Score against neighborhood width for three lead days. At one-day lead the score rises steeply from 0.33 at 0.1° to 0.50 at 0.3° in 2023 and from 0.42 to 0.63 in 2024, then flattens beyond about 1°; each curve crosses its useful-skill threshold (0.506 and 0.511) at a width of approximately 0.34° in 2023 and 0.19° in 2024 — that is, at two to three grid cells, or roughly 20–40 km. Most of the displacement error is therefore contained within a few pixels rather than representing a gross misplacement of fire activity.

This scale is relevant to the intended application. GEFS-Aerosols runs on a C384 grid with an effective grid spacing of about 25 km, and NOAA's next-generation global aerosol/chemistry systems based on the Unified Forecast System (UFS), such as the Global Chemistry and Aerosol Forecast System (GCAFS) and future ones based on forthcoming chemistry components (UFS-Chem) are also at comparable resolution; both ingest emissions interpolated from the 0.1° FRP field rather than resolving that field directly. The model's effective resolution is therefore close to the scale at which the receiving model can represent the emitted plume, even though the emission fields are supplied on the finer grid.

The behavior with lead time differs markedly between the two years. In 2023 the effective resolution coarsens substantially as the forecast extends, from about 0.3° at D+1 to roughly 2° at D+7, so that by the end of the window useful skill is attained only after aggregation over a large neighborhood. In 2024 all three lead days cross the threshold near 0.2°, and the curves remain separated by less than 0.05 in FSS across the whole range of scales. The model's spatial placement in 2024 is thus not only better at short lead but also far more persistent across the forecast window — a difference consistent with the smaller decline in categorical skill in that year, though its cause has not been isolated.

*5.3. Skill Relative to Operational Practice*

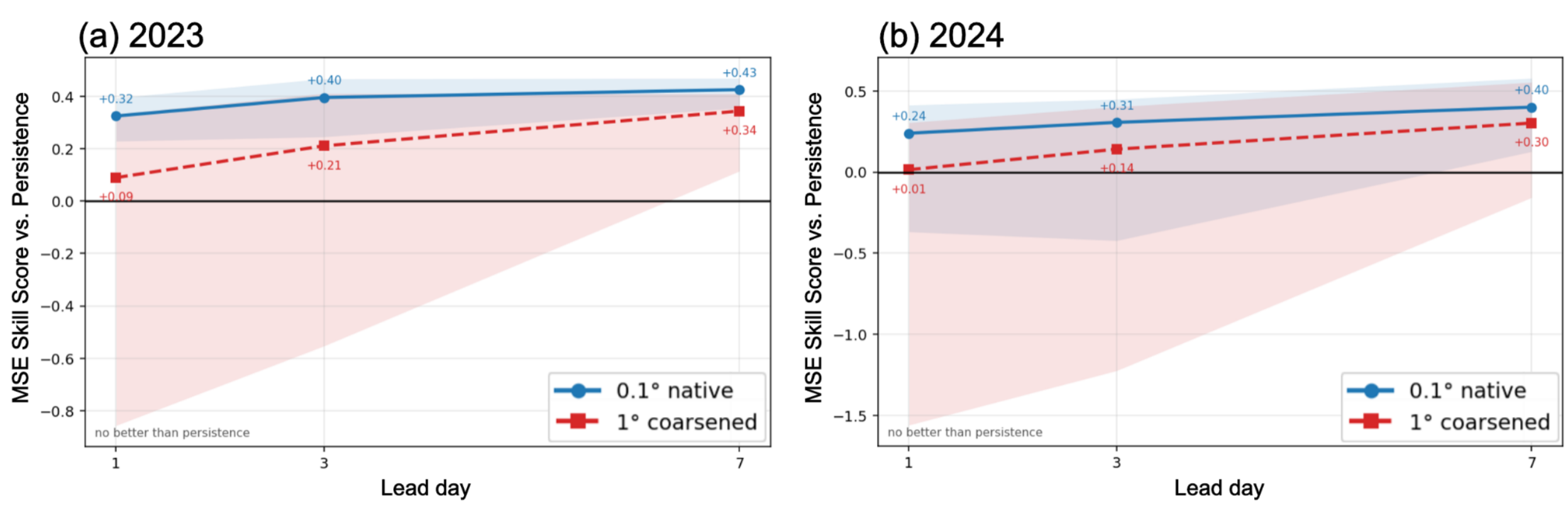

**Figure 9.** MSE skill score relative to persistence of the initialization-time observed FRP, held fixed across the forecast as in operational practice. (a) 2023; (b) 2024. Blue, 0.1° native; red, 1° aggregated. Shading spans the twelve monthly values; the horizontal line marks equivalence.

Figure 9 shows the mean-squared-error skill score against that reference, defined as 1 − MSE_model / MSE_persistence, for which positive values indicate lower error than the persisted field. The 0.1° native resolution is the operationally relevant one, since GBBEPx is produced on that grid and the emission fields are supplied to GEFS-Aerosols on it; the 1° values are reported alongside to show how the comparison behaves once small-scale displacement is averaged out. The aerosol model itself runs on a C384 grid of roughly 25 km, so the scales it resolves lie between the two, closer to the native grid. At 0.1° the model reduces mean squared error by 32% at one-day lead, 40% at three days and 43% at seven days in 2023, and by 24%, 31% and 40% in 2024. At 1° the corresponding reductions are 9%, 21% and 34% in 2023, and 1%, 14% and 30% in 2024.

The improvement relative to persistence is larger at the native resolution than after aggregation, because the persisted field degrades faster there: a fire front that has moved by a single pixel is entirely misplaced in the persisted analysis, whereas the model still reproduces its approximate distribution. The absolute scores at native resolution are weaker than at 1°, but the margin over current operational practice is wider — and it is at 0.1° that the comparison matters.

The model's own skill declines with lead time, but the persisted field degrades faster, so the margin between them widens. At D+1 the most recent observation is still a strong predictor, and the model adds relatively little at the aggregated scale; by D+7 the persisted field is a week out of date and the model reduces its error by a third or more. The advantage is largest precisely where the operational assumption is weakest — in the later days of the forecast window, which are also the days for which a predictive capability is most needed.

A day-of-year climatological reference was also examined but is not reported. The available climatology requires a minimum number of historical fire days at a given pixel and calendar day, and is consequently undefined for 99.8% of pixel–day combinations; after gap-filling it represents only about a fifth of the observed variance, and its low mean squared error reflects the variance term of the error decomposition rather than forecast accuracy. Persistence, whose variance matches the observations by construction, and which is also the reference used operationally, is the appropriate comparison here.

## 6. Summary and Discussion

The model addresses the sparse, heavy-tailed statistics of fire activity through a hurdle prediction head and fire-tailored composite loss, is trained separately over five fire-regime domains, and applies a per-pixel intensity cap anchored to historical FRP climatology.

Verified against two independent years, 2023 and 2024, the model reproduces the global seasonal cycle of daily FRP at both resolutions, with median absolute daily global-mean bias below 0.05 MW per 0.1° pixel at the shortest lead in both years. Relative to persistence of the most recent observed field—the reference implied by current operational practice—it reduces mean squared error by roughly a quarter to a third at one-day lead and by around 40% at seven days at 0.1° resolution. The advantage increases with lead time as the persisted field becomes increasingly outdated.

The clearest result is the asymmetry between detection and intensity skill. Across the seven-day forecast window, ETS declines by only a few hundredths, whereas the Kling–Gupta efficiency of predicted intensity declines by an order of magnitude more in every month of both years. Detection probability increases with observed fire intensity, indicating that large fires are located more reliably than small ones. The model therefore remains effective at identifying which pixels are burning but substantially underestimates how strongly they burn, predicting only about a quarter to a third of observed intensity in the largest-fire stratum. The climatology-constrained cap is not responsible; the scarcity of extreme fires in the training distribution and the clipping of FRP inputs and targets are more plausible causes.

Skill is consistently higher in 2024 than in 2023 across nearly every metric, despite 2024 being farther from the training period; the cause remains unresolved. Native-resolution false alarms are also primarily spatial-displacement errors rather than weak spurious predictions: increasing the detection threshold enough to remove much of their radiative power would also eliminate many correctly located fires. The Fractions Skill Score shows an effective resolution of about two to three grid cells at one-day lead, substantially finer than suggested by the native 0.1° categorical scores, although this resolution coarsens with lead time in 2023.

Absolute-error metrics are dominated by the largest fires, so regions or months with similar relative performance can exhibit very different MAE or RMSE depending on their intensity distributions. Skill is also spatially heterogeneous. Siberia shows substantially poorer boreal-season detection than North America despite both being boreal forest regions and using the same architecture and loss. Siberia lies at the northern margin of the Southeast Asia training domain, which spans 10°S–85°N and combines fire regimes ranging from equatorial peatland burning to Arctic tundra. This domain required a fiftyfold reduction in learning rate to converge, likely limiting learning at its high-latitude margin. We therefore attribute the Siberian weakness primarily to excessive heterogeneity within the training domain rather than to the GNN architecture itself.

The principal remaining limitation is intensity calibration. The largest-fire stratum is underestimated by a factor of roughly three to four, and spatial aggregation does not recover the missing radiative power. Quantile-based output heads, upper-tail correction, or ensemble approaches are therefore priorities for improving emission forcing. Spatial displacement is a second concern, although GEFS-Aerosols operates on a C384 grid of roughly 25 km, comparable to the model's effective short-lead resolution. The current two-layer, four-neighbor architecture has a receptive field of only ±2 cells (0.2°), similar to the scale identified by the FSS analysis, suggesting that greater network depth or connectivity may improve spatial skill. A spatially tolerant loss, such as a differentiable Fractions Skill Score, could also reduce the double penalty imposed when a fire is displaced by one grid cell, while higher-resolution fuel, terrain, and anthropogenic ignition predictors may provide additional spatial information.

The lead-dependent damping of variability motivates stochastic perturbation or lagged-ensemble methods, and the training domain boundaries should be revised so that high-latitude regions are not sampled against tropical fire regimes, particularly by splitting Southeast Asia by latitude. The regional design was adopted because of memory and training-time constraints at 0.1° and because fire regimes and FRP distributions differ strongly between regions; a single global model, trained jointly with region-aware inputs, is a complementary direction to test whether shared learning across regimes improves generalization. Finally, the verification here uses reanalysis meteorology as predictor input; in operational use, the predictors would themselves be forecast quantities, so the reported skill should be read as an upper bound, and quantifying that reduction is an immediate priority.

Converting predicted FRP into biomass-burning emissions is a separate step, introducing uncertainty from combustion completeness and emission factors that is independent of FRP forecast error; supplying the converted fields to UFS-based air quality models and assessing their effects on forecasts of aerosol optical depth, surface PM2.5, and plume transport is the ultimate test of operational value and the focus of subsequent work.

## Acknowledgments

This research was supported in part by the Bipartisan Infrastructure Law (BIL) NA23OAR4050201I, NOAA Weather Program Office (https://ror.org/01q70bz83) under award NA23OAR4590387, and NOAA cooperative agreement NA22OAR4320151, for the Cooperative Institute for Earth System Research and Data Science (CIESRDS).

## Data availability statement.

The GBBEPx v4 fire radiative power data used for model training and evaluation are produced by NOAA/NESDIS and are available from https://www.ospo.noaa.gov/pub/Blended/GBBEPx/GBBEPx-all01GRID (last accessed: 22 September 2026). The duration of the fire predictor was derived from the same GBBEPx fire radiative power record and requires no additional data source. The ERA5 reanalysis dataset (0.25° grid), from which the daily meteorological predictors were derived, was obtained from the NSF NCAR GDEX repository (https://doi.org/10.5065/BH6N-5N20, last accessed: 22 September 2026), which contains modified Copernicus Climate Change Service information. Vegetation Health Index data are available from NOAA/NESDIS (https://www.star.nesdis.noaa.gov/smcd/emb/vci/VH/index.php; last accessed: 22 September 2026), and land-cover data are available from NASA's EarthData system (https://doi.org/10.5067/MODIS/MCD12C1.061). The model code is available at https://doi.org/10.5281/zenodo.23196414.

## Disclaimer

The statements, findings, conclusions, and recommendations are those of the author(s) and do not necessarily reflect the views of NOAA or the U.S. Department of Commerce.

## APPENDIX

### Spatial Evaluation of Predicted Fire Radiative Power

To further illustrate the spatial performance of ML-FIRE, monthly mean fire radiative power (FRP) is compared with GBBEPx observations for representative months in 2023 and 2024. Figures A1 and A2 show results at one- and seven-day forecast lead times, respectively, at both the native 0.1° resolution and a coarsened 1° resolution.

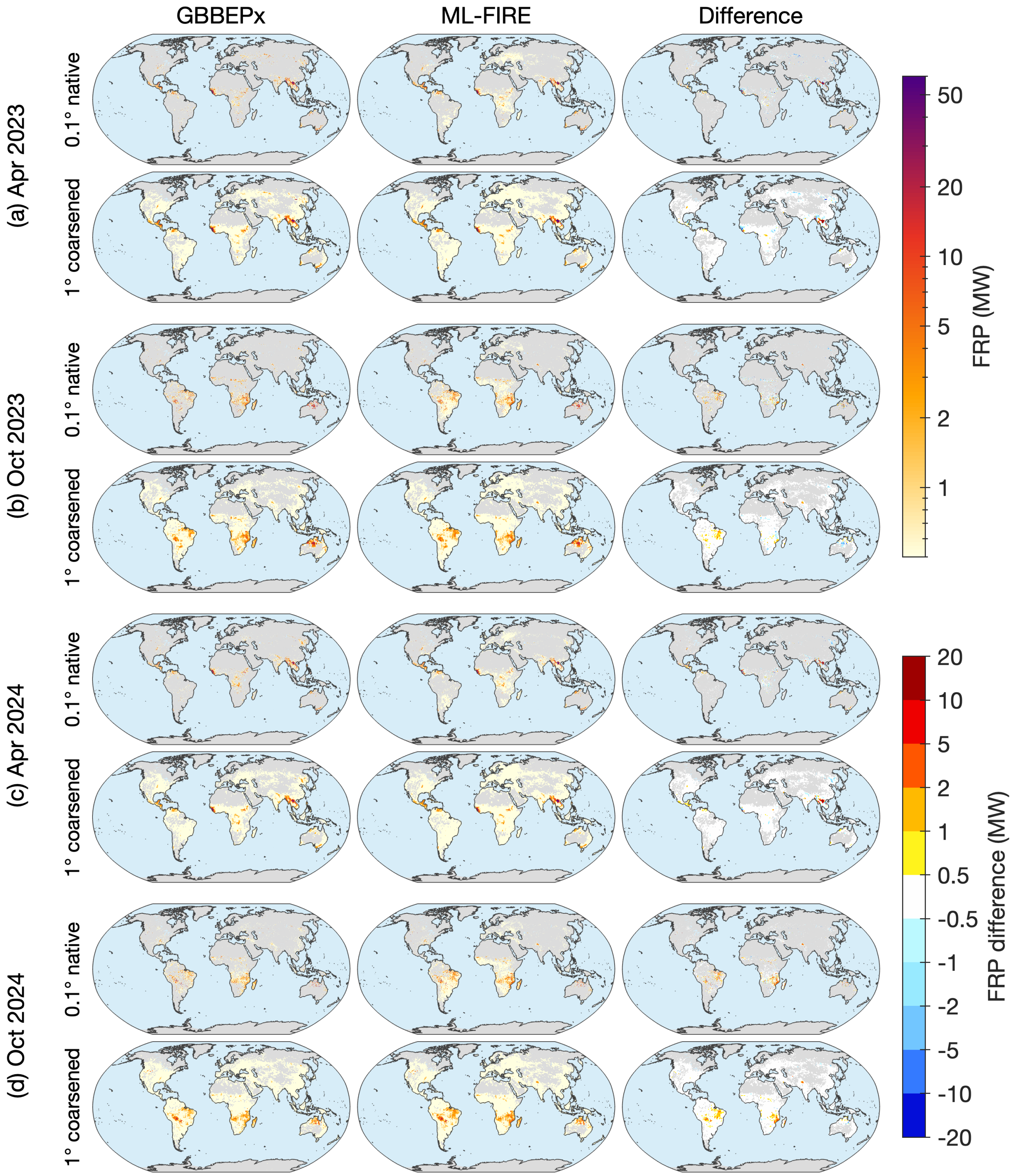

**Figure A1.** Monthly mean FRP at a one-day forecast lead time for (a) April 2023, (b) October 2023, (c) April 2024, and (d) October 2024. Columns show GBBEPx, ML-FIRE predictions, and their difference (ML-FIRE − GBBEPx). Within each panel, the upper and lower rows show the native 0.1° and coarsened 1° resolutions, respectively. Color scales denote FRP and FRP differences.

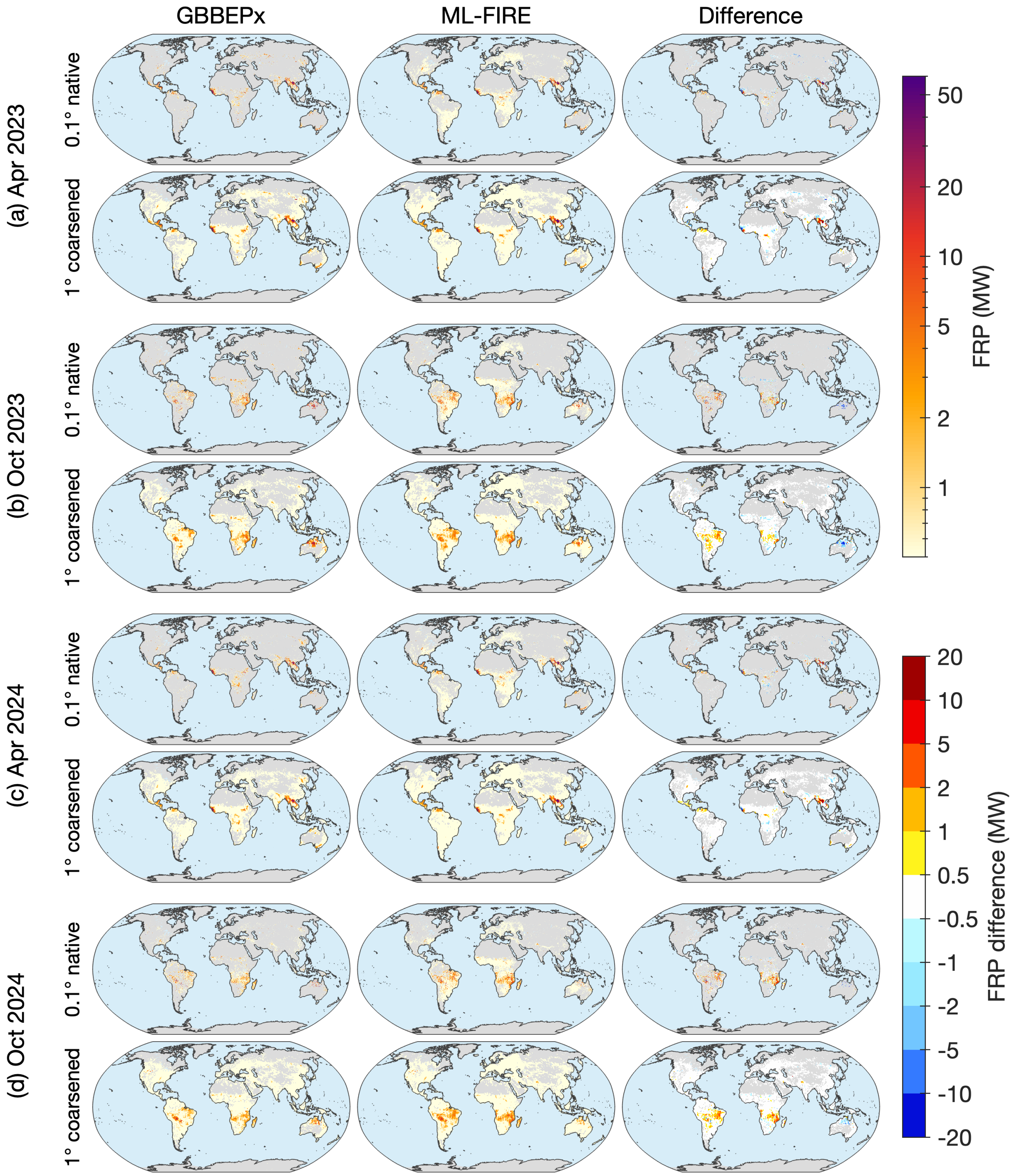

**Figure A2.** As in Fig. A1, but for a seven-day forecast lead time.